\documentstyle[prl,aps]{revtex}
\newcommand{\bb}[1]{\mbox{\boldmath ${#1}$}}
\begin{document}
%------------------------------------------------------------------------------
\title{From the Liouville Equation to the Generalized Boltzmann Equation for
Magnetotransport in the 2D Lorentz Model}
\author{A.\ V.\ Bobylev\footnote{Permanent address: Keldysh Institute for
Applied Mathematics, Russian Academy of Science, 125047 Moscow, Russia}}
\address{Department of Physics, Norwegian University of Science and
Technology, N--7491 Trondheim, Norway}
\author{Alex Hansen\footnote{Permanent address: Department of Physics, 
Norwegian University of Science and Technology, N--7491 Trondheim, Norway}}
\address{Laboratoire de Physique de l'Ecole Normale Sup\'erieure de Lyon,
F--69364 Lyon Cedex 07, France}
\author{J.\ Piasecki}
\address{Institute of Theoretical
Physics, University of Warsaw, ul.\ Ho{\.z}a 69, PL--00 681 Warsaw, Poland}
\author{E.\ H.\ Hauge}
\address{Department of Physics, 
Norwegian University of Science and Technology, N--7491 Trondheim, Norway}
\date{\today}
\maketitle
%------------------------------------------------------------------------------
\begin{abstract} 
We consider a system of non-interacting charged particles moving in two
dimensions among fixed hard scatterers, and acted upon by a perpendicular 
magnetic field.  Recollisions between charged particles and scatterers are 
unavoidable in this case.  We derive from the Liouville equation for this 
system a generalized Boltzmann equation with infinitely long memory, but 
which still is analytically solvable.  This kinetic equation has been earlier 
written down from intuitive arguments.
\end{abstract} 
%------------------------------------------------------------------------------
% MAIN TEXT
%------------------------------------------------------------------------------
\section{Introduction}
\label{intro}
The Lorentz model \cite{L05,HS,H74} where a particle moves among and collides 
with fixed scatterers has provided a rich testing ground for kinetic  
theory.  In particular, the Boltzmann equation is not only exactly solvable 
in this model, but the equation itself is exact in the Grad limit (to be
defined below).  The {\it Stosszahlansatz} used in constructing the Boltzmann 
equation relies implicitly on 
the assumption that the particle never returns to a scatterer
after having collided with it. The probability of such recollisions vanishes in
the Grad limit. On this basis the Boltzmann equation for the standard
Lorentz model is taken to
be exact in the Grad limit. 

Recently, however, an interesting exception was discovered \cite{BMHH95,BMHH97} 
to this state of affairs.
Let the particle have electric charge $-e$ and move on 
a plane pierced by a perpendicular constant magnetic field $B$.  The particle
moves along circle arcs between collisions, and if it does not 
encounter a new scatterer along the arc, it will 
{\it recollide\/} with the initial scatterer.  This destroys the 
above assumption and renders the Boltzmann equation invalid. Note that
both the (two-)dimensionality of space and the presence of a perpendicular
magnetic field are essential here.

In Refs.\ \cite{BMHH95,BMHH97} a generalization of the conventional
Boltzmann equation was proposed that takes into account 
consecutive recollisions with the same scatterer.  The arguments leading
to this equation were intuitive, on the same level as the
{\it Stosszahlansatz\/} itself. The aim of the present
paper is to provide a microscopic underpinning to the generalized
Boltzmann equation in the form of
a systematic derivation from the Liouville equation.
As a background we start by summarizing the intuitive arguments leading to 
the generalized Boltzmann equation.

The charged particle moves on a plane of (large) area $A$, with 
$N$ randomly placed hard disk scatterers of radius $a$.  We denote by $n=N/A$ 
their number density. The disks do not overlap.

The generalized Boltzmann equation describes the evolution of the 
probability density $f_1({\bf x},{\bf v},t)$ for finding the moving
particle at time $t$ at position $\bf x$ with velocity $\bf v$. 
This non-markovian kinetic equation has the form
\begin{equation}
\label{fg1}
\frac{D}{Dt}\ \ f^G({\bf x},{\bf v},t)=
na\ \sum_{k=0}^{[t/T_0]}\ e^{-\nu kT_0}\ 
\int_{S^1} d{\bf n} ({\bf v}\cdot {\bf n})
[\theta({\bf v}\cdot {\bf n})b_{\bf n} +\theta(-{\bf v}\cdot {\bf n})]\
f^G\left({\bf x},S_0^{-k}{\bf v},t-kT_0\right)\;,
\end{equation}
where
\begin{equation}
\label{fgdef}
f^G({\bf x},{\bf v},t)=
\left\{ \begin{array}{rl}
      f_1({\bf x},{\bf v},t) & \mbox{if $0<t<T_0$}\\
      (1-e^{-\nu T_0}) f_1({\bf x},{\bf v},t) & \mbox{if $t>T_0$}\;,
      \end{array}
      \right.
\end{equation}
and where
$\nu =2|{\bf v}|na$ is the collision frequency and $T_0$ the cyclotron period.
Furthermore,
\begin{equation}
\label{totaldt}
\frac{D}{Dt}=
\left[\frac{\partial}{\partial t} + {\bf v} \cdot \frac{\partial}{\partial
{\bf x}} +\left( \bb{\omega}\times{\bf v}\right)\cdot
\frac{\partial}{\partial {\bf v}}\right]
\end{equation}
is the generator of free cyclotron motion with frequency $|\bb{\omega}|
=\omega=eB/m$, and $[t/T_0]$ the number of cyclotron periods 
$T_0=2\pi/\omega$ completed before time $t$.
The angular integration over the unit vector {\bf n} 
in (\ref{fg1}) is over the entire unit sphere $S^1$
centered at the origin.  In the gain term (positive contribution), there 
appears the distribution acted upon by the operator $b_{\bf n}$, defined by
\begin{equation}
\label{gain}
b_{\bf n} \phi({\bf v})=\phi({\bf v}-2({\bf v}\cdot{\bf n}){\bf n})\;,
\end{equation}
where $\phi$ is an arbitrary  function of $\bf v$.
The precollisional velocity ${\bf v}'={\bf v}-2({\bf v}\cdot{\bf n}){\bf n}$
becomes $\bf v$ after the elastic collision with the immobile (infinitely
massive) scattering disk.  Note that ${\bf v}'\cdot{\bf n}<0$.  In the 
loss term (negative contribution), the precollisional velocity, $\bf v$,
is also from the hemisphere ${\bf v}\cdot {\bf n}<0$. Finally, the shift
operator $S_0^{-k}$, when acting on $\bf v$, rotates the velocity through
the angle $-k\psi$, where $\psi$ is the scattering angle (from $\bf v'$
to $\bf v$). 

When $0<t<T_0$, $[t/T_0]=0$, and $f^G({\bf x},{\bf v},t)=
f_1({\bf x},{\bf v},t)$. No recollisions are yet possible and 
equation (\ref{fg1}) reduces to the standard
Boltzmann equation
\begin{equation}
\label{normalbolt}
\frac{D}{Dt}\ \ f_1({\bf x},{\bf v},t)=
na\ \int_{S^1} d{\bf n} ({\bf v}\cdot {\bf n})
[\theta({\bf v}\cdot {\bf n})b_{\bf n} +\theta(-{\bf v}\cdot {\bf n})]\
f_1\left({\bf x},{\bf v},t\right)={\cal C}^B f_1({\bf x},{\bf v},t)\;,
\end{equation}
where ${\cal C}^B$ is the Boltzmann collision operator.

When $t>T_0$, the distribution $f_1({\bf x},{\bf v},t)$ splits into two parts.
With probability $\exp(-\nu T_0)$ the charged
particle continues to perform free cyclotron motion, having explored
the whole circle during the first period $T_0$.  With probability 
$[1-\exp(-\nu T_0)]$ the particle suffers collisions and becomes a
wandering particle among the hard disks. Since the probability of being
a wandering particle is less than unity, 
the distribution for $t>T_0$ should be renormalized. Hence
the need for (\ref{fgdef}).

The sum in the generalized equation (\ref{fg1}), from $k=1$ to $k=[t/T_0]$, 
takes into account all 
possible recollision events. The collision ``now" can be the
$k$'th recollision, with  every recollision having the same
scattering angle $\psi$, equal to the one of 
the initial collision at $t-kT_0$. Only this initial collision, with
incoming velocity $S_0^{-k}{\bf v}$,
is described by the Boltzmann operator ${\cal C}^B$, all subsequent ones
follow from dynamics alone, weighted by the survival probability
from one collision to the next, $\exp (-\nu T_0)$. Thus, for a
sequence of $k$ recollisions on gets the factor $\exp (-k\nu T_0)$. 
Finally, summation over 
$k$ from 1 to $[t/T_0]$ takes all possible recollision sequences into account.

On the basis of an intuitive derivation, as sketched above, and by analogy
to the results for the general Lorentz model, Refs.\ \cite{BMHH95,BMHH97} 
assume that the generalized Boltzmann equation (\ref{fg1}) gives an exact description
of the time evolution of our system in the Grad limit,
\begin{equation}
\label{graddef}
\lim_{\rm Grad}=
\left\{ \begin{array}{l}
      a\to 0\quad\mbox{and}\quad N/A=n\to\infty \\
      na=\mbox{const}\;.
      \end{array}
      \right.
\end{equation}
According to (\ref{graddef}) the radius of the scattering disks approaches
zero but their number density increases at the same time in such a way that
the mean free path of the wandering particle, $\Lambda =(2na)^{-1}$ remains constant.

In the next section, we discuss the initial value problem for the Liouville
equation. In Section \ref{recollisions}, the effect of recollisions is
analyzed, and Section \ref{genboltz} contains the derivation of the generalized 
Boltzmann equation.  We conclude with a short summary. The shift operator 
along the trajectory involving recollisions is described in the appendix.
  
%------------------------------------------------------------------------------
\section{Dynamical Equations}
\label{dyneq}
We denote by $F({\bf x},{\bf v},t;{\bf y}_1,...,{\bf y}_N)$ the joint
probability density for finding the moving point charge at time $t$ at
point $\bf x$ with velocity $\bf v$ among $N$ hard disk scatterers located
at positions ${\bf y}_1$, ${\bf y}_2$,...,${\bf y}_N$.  $F$ satisfies
the normalization condition
\begin{equation}
\label{normaliz}
\int_{\bf R^2} d{\bf v} \int_{\bf \Omega} d{\bf x} 
\int_{\Omega^N} d{\bf y}_1...d{\bf y}_N 
F({\bf x},{\bf v},t;{\bf y}_1,{\bf y}_2,...,{\bf y}_N)=1\;,
\end{equation}
and is the solution of the Liouville equation
\begin{equation}
\label{liouville}
\frac{D}{Dt} F\left({\bf x},{\bf v},t;{\bf y}_1,{\bf y}_2,...,{\bf y}_N\right)=
a\sum_{j=1}^N T({\bf x}-{\bf y}_j,{\bf v}) 
F({\bf x},{\bf v},t;{\bf y}_1,{\bf y}_2,...,{\bf y}_N)\;,
\end{equation}
where $T({\bf x}-{\bf y}_j,{\bf v})$ is the binary collision operator for
scatterer $j$,
\begin{equation}
\label{coll}
T({\bf x}-{\bf y}_j,{\bf v})=
\int_{S^1} d{\bf n} \left({\bf v} \cdot {\bf n}\right)
\left[\theta({\bf v} \cdot {\bf n})b_{\bf n}
+\theta(-{\bf v} \cdot {\bf n})\right] \delta({\bf x}-{\bf y}_j -a
{\bf n})\;.
\end{equation}
The operator $b_{\bf n}$ acting on the velocity vector has been defined in 
(\ref{gain}), and $a{\bf n}$ is the point of impact with respect to the 
center of the
disk.

At time $t=0$, the initial condition for the Liouville equation 
(\ref{liouville}) is assumed to be of the form
\begin{equation}
\label{initialcond}
F({\bf x},{\bf v},t=0;{\bf y}_1,...,{\bf y}_N)
=f_1({\bf x},{\bf v},0)\left[\prod_{i=1}^N \ 
\theta(|{\bf x}-{\bf y}_i|-a)\right]\ 
\frac{\rho({\bf y}_1,...,{\bf y}_N)}{1-\pi n a^2}\;.
\end{equation}
Here $f_1({\bf x},{\bf v},0)$ is the initial state of the particle, and
$\rho({\bf y}_1,...,{\bf y}_N)$ describes the static, non-overlapping 
distribution of scattering disks,
\begin{equation}
\label{rho}
\rho({\bf y}_1,...,{\bf y}_N)
=\prod_{1\le i<j\le N} \theta(|{\bf y}_i-{\bf y_j}|-2a)/{\cal N}\;,
\end{equation}
where ${\cal N}$ is the normalizing factor such that $\int_{\Omega^N}
d{\bf y}_1\cdots{\bf y}_N\rho({\bf y}_1,...,{\bf y}_N)=1$. 

To complete the statement of the problem, we shall for simplicity assume that 
the region $\Omega$ enclosing the system has finite area $A$ and no 
boundaries (e.g., a two-dimensional torus).  We furthermore assume that the
linear dimensions of $\Omega$ are very large compared to the cyclotron
radius. 

The factor $[1-\pi n a^2]^{-1}$ assures the proper normalization 
(\ref{normaliz}) of the initial condition since 
\begin{equation}
\label{rhonorm}
\int_{\Omega^N} d{\bf y}_1...d{\bf y}_N  
\rho({\bf y}_1,...,{\bf y}_N)\
\left[\prod_{i=1}^N \ \theta(|{\bf x}-{\bf y}_i|-a)\right]\
=1-\pi n a^2\;.
\end{equation}

Clearly, the probability density $f_1({\bf x},{\bf v},t)$ is obtained
from $F({\bf x},{\bf v},t;{\bf y}_1,...,{\bf y}_N)$ by integration over
the positions of all scatterers,
\begin{equation}
\label{ftof1}
f_1({\bf x},{\bf v},t)=
\int_{\Omega^N} d{\bf y}_1...d{\bf y}_N  
F({\bf x},{\bf v},t;{\bf y}_1,...,{\bf y}_N)\;.
\end{equation}
Integrating the Liouville equation (\ref{liouville}) we thus find
\begin{equation}
\label{hier5}
\frac{D}{Dt}f_1({\bf x},{\bf v},t)=aN\int_\Omega d{\bf y}_1\ 
T({\bf x}-{\bf y}_1,{\bf v}) f_2({\bf x},{\bf v},t;{\bf y}_1)\;,
\end{equation}
where
\begin{equation}
\label{ftof2}
f_2({\bf x},{\bf v},t;{\bf y}_1)=
\int_{\Omega^{N-1}} d{\bf y}_2...d{\bf y}_N  
F({\bf x},{\bf v},t;{\bf y}_1,...,{\bf y}_N)\;.
\end{equation}
Equation (\ref{hier5}) will be the main object of our study.  Using the
explicit form of the collision operator (\ref{coll}), we rewrite it as
\begin{equation}
\label{basic}
\frac{D}{Dt}f_1({\bf x},{\bf v},t)=
an\ \int_{S^1} d{\bf n} ({\bf v}\cdot {\bf n}) 
[\theta({\bf v}\cdot {\bf n})b_{\bf n} +\theta(-{\bf v}\cdot {\bf n})]\
I_a({\bf x},{\bf v},{\bf n},t)\;,
\end{equation}
where
\begin{equation}
\label{ia}
I_a({\bf x},{\bf v},{\bf n},t)
=A\ \int_{\Omega^{N-1}} d{\bf Y} F({\bf x},{\bf v},t;
{\bf x}-a{\bf n},{\bf Y})\;.
\end{equation}
where from now on we use the short-hand notation ${\bf Y}=
({\bf y}_2,...{\bf y}_N)$.

In Eq.\ (\ref{ia}), ${\bf y}_1={\bf x}-a{\bf n}$, which corresponds to
the precollisional presence of the particle at the surface of the scatterer.
As has already been mentioned in the Introduction, only velocities such
that $({\bf v}\cdot{\bf n})<0$ appear as argument in the distribution
$F$.

Our aim here is to demonstrate that Eq.\ (\ref{basic}) becomes a
closed kinetic equation for the distribution $f_1({\bf x},{\bf v},t)$ in
the Grad limit (\ref{graddef}), and that taking this limit we shall
recover the generalized Boltzmann equation (\ref{fg1}).

We note that the Liouville equation (\ref{liouville}) implies the
relation
\begin{equation}
\label{formalsol}
F({\bf x}={\bf y}+a{\bf n},{\bf v},t;{\bf y},{\bf Y})
=F(S^{(N)}_{-t} ({\bf y}+a{\bf n}),S^{(N)}_{-t}{\bf v},t=0;
{\bf y},{\bf Y})\;
\end{equation}
where $S^{(N)}_{-t}$ denotes the backwards shift operator along the exact phase
trajectory of the $N$-scatterer problem.  A crucial point is to
separate clearly two possibilities: (a) The particle did not collide with the
scatterer at $\bf y$ before time $t$. (b) It did (leading to the problem of
recollisions). The concequences of this distinction will be discussed 
in the next section.
%------------------------------------------------------------------------------
\section{Recollisions}
\label{recollisions}
For any fixed configuration of scatterers $({\bf y},{\bf Y})
=({\bf y},{\bf y}_2,...,{\bf y}_N)$, one can construct the past history for 
the moving particle leading to the position ${\bf y}+a{\bf n}$ and velocity 
${\bf v}$ at time $t$.  The probability weight for the initial conditions at 
$t=0$ is defined in Eq.\ (\ref{initialcond}). 

At $t>0$, when ${\bf x}={\bf y}+a{\bf n}$, the particle is about to collide with
the scatterer at $\bf y$, since $({\bf v}\cdot{\bf n})<0$.  Depending on the 
past history, we separate
the integration domain $\Omega^{N-1}$ in (\ref{ia}) into a union of 
disjoint time-dependent subdomains $\Delta_k({\bf y}+a{\bf n},{\bf v},t)$ 
defined as
\begin{equation}
\label{delta0}
\Delta_0({\bf y}+a{\bf n},{\bf v},t)
=\{{\bf Y}\ :\ \mbox{no collision with scatterer at $\bf y$ 
before $t$}\}\;,
\end{equation}
and 
\begin{equation}
\label{deltak}
\Delta_k({\bf y}+a{\bf n},{\bf v},t)
=\{{\bf Y}\ :\ \mbox{$k$ collisions with scatterer at $\bf y$ before
$t$}\}\;,
\end{equation}
where $k=1,2,...,\infty$.  Clearly, we have
\begin{equation}
\label{snitt}
\Delta_j({\bf y}+a{\bf n},{\bf v},t)\cap
\Delta_k({\bf y}+a{\bf n},{\bf v},t)=\emptyset
\end{equation}
when $j\neq k$, and
\begin{equation}
\label{union}
\bigcup_{k=0}^\infty \Delta_k({\bf y}+a{\bf n},{\bf v},t)=\Omega^{N-1}\;.
\end{equation}
Thus, we may rewrite Eq.\ (\ref{ia}) as
\begin{equation}
\label{ia1}
I_a({\bf y}+a{\bf n},{\bf v},{\bf n},t)
=\sum_{k=0}^\infty A\ 
\int_{\Delta_k({\bf y}+a{\bf n},{\bf v},t)} 
d{\bf Y}\ F({\bf y}+a{\bf n},{\bf v},t;
{\bf y},{\bf Y})\;.
\end{equation}

We now introduce a further partition of each
 $\Delta_k({\bf y}+a{\bf n},{\bf v},t)$, 
$k=1,..,\infty$, into two disjoint parts,
\begin{equation}
\label{disjoint}
\Delta_k({\bf y}+a{\bf n},{\bf v},t)
=\Delta_k^{(0)}({\bf y}+a{\bf n},{\bf v},t)
\cup\Delta_k^{(1)}({\bf y}+a{\bf n},{\bf v},t)\;,
\end{equation}
\begin{equation}
\label{junion}
\Delta_k^{(0)}({\bf y}+a{\bf n},{\bf v},t)
\cap\Delta_k^{(1)}({\bf y}+a{\bf n},{\bf v},t)
=\emptyset ,
\end{equation}
such that
\begin{equation}
\label{deltak0}
\Delta_k^{(0)}({\bf y}+a{\bf n},{\bf v},t)
=\{{\bf Y}\ :\ \mbox{no collisions with scatterers at $\bf Y$ within the time
interval $[t-kT_a,t]$}\}\;.
\end{equation}
Here $T_a$ is the period between two successive collisions of the particle
with the scatterer at $\bf y$.

If ${\bf Y}\in \Delta_k^{(0)}({\bf y}+a{\bf n},{\bf v},t)$, then the particle
collided with the scatterer at $\bf y$ $k$ times during the time interval
$[t-kT_a,t]$.  Furthermore, it had not collided with this scatterer before
$t-kT_a$.  We may therefore conclude that 
\begin{equation}
\label{yin0}
{\bf Y}\in\Delta_k^{(0)}({\bf y}+a{\bf n},{\bf v},t)\ \Rightarrow\
F({\bf y}+a{\bf n},{\bf v},t;{\bf y},{\bf Y})=
F(S_a^{-k}({\bf y}+a{\bf n}),S_a^{-k}{\bf v},t-kT_a;{\bf y},{\bf Y})
\end{equation}
where $S_a$ is the shift operator along the trajectory of the particle
between two collisions with the sactterer at $\bf y$, in the absence 
of other scatterers.
We construct this operator explicitly in the Appendix, see also \cite{BMHH97}.

In the Grad limit,
the volume of $\bf Y$-space for which collisions with other
scatterers occur between recollisions with the same scatterer is
negligible compared to the volume corresponding to a sequence of consecutive 
recollisions, and in fact
\begin{equation}
\label{fundamentaleq}
\lim_{\rm Grad}\|\Delta^{(1)}_k\| =0\;,
\end{equation}
where $\|\cdots\|$ represents the relative volume with respect to the
set $\Omega^{N-1}$.
The reason behind Eq.\ (\ref{fundamentaleq})
is that the probability to return to the scatterer at 
$\bf y$ after an intermediate collision with another scatterer involves an 
extra power of $a$ and thus vanishes in the Grad limit. The corresponding 
problem in the absence of a magnetic field is discussed in \cite{HS}.
Close to the Grad limit, (\ref{fundamentaleq}) allows us to change the
volume of integration in (\ref{ia1}) from $\Delta_k({\bf y}+a{\bf n},{\bf v},t)$
to $\Delta_k^{(0)}({\bf y}+a{\bf n},{\bf v},t)$.  

Let us introduce the limit function
\begin{equation}
\label{gradf}
{\cal F}_k({\bf y},S_0^{-k}{\bf v},t-kT_0) =\lim_{\rm Grad}
\left(\frac{A}{\|\Delta_k^{(0)}({\bf y}+a{\bf n},{\bf v},t)\|}\right)\ 
\int_{\Delta_k^{(0)}({\bf y}+a{\bf n},{\bf v},t)}
d{\bf Y}\ F({\bf y}+a{\bf n},{\bf v},t;{\bf y},{\bf Y})\; ;\quad
\Delta_0^{(0)}\equiv \Delta_0\;.
\end{equation}
where (\ref{yin0}) has been used for convenience. This limit function 
no longer depends on the number of scatterers, $N$.
 
In terms of ${\cal F}_k$, the Grad limit of (\ref{ia1}) becomes
\begin{equation}
\label{ia2}
I_0({\bf y},{\bf v},{\bf n},t)=\sum_{k=0}^{[t/T_0]} 
\|\Delta_k^{(0)}({\bf y},{\bf v},t)\|\
{\cal F}_k ({\bf y},S_0^{-k}{\bf v},t-kT_0)\;.
\end{equation}
Note, with reference to the Appendix,  that the shift operator 
$S_0^{-k}$ depends on the vector $\bf n$, even 
in the Grad limit.

The relation
\begin{equation}
\label{deltaequal}
\Delta_k^{(0)}({\bf y},{\bf v},t)=\Delta_0
({\bf y},S_0^{-k}{\bf v},t-kT_0)\cap \Gamma(t-kT_0,t)\;,
\end{equation}
is valid in the Grad limit where the set $\Gamma(t_1,t_2)$ corresponds to 
those regions of $\bf Y$-space in which the particle suffers 
no collisions with the $N-1$ scatterers at $\bf Y$ during the time interval 
$(t_1,t_2)$.  

We now evaluate the volumes of the sets $\Delta_0
({\bf y},S_0^{-k}{\bf v},t-kT_0)$ and $\Gamma(t_1,t_2)$. In order to do so, 
we define the set complementary to $\Delta_0({\bf y},{\bf v},t)$,
\begin{equation}
\label{deltabar}
\overline{\Delta}_0({\bf y},{\bf v},t)
=\{{\bf Y}\ :\ \mbox{at least one 
collision with scatterer at $\bf y$ before $t$}\}\;,
\end{equation}
If $t\geq T_0$, two types of histories consistent with the condition in
(\ref{deltabar}) are possible: (a) No collisions with other scatterers
occured within the time interval $(t-T_0,t)$. (b) The particle collided
with other scatterers after the previous collision with the scatterer
at $\bf y$. Close to the Grad limit possibility (b) is improbable and
in the limit, possibility (a) is realized with probability 1. If, on the
other hand, $t<T_0$, the condition in (\ref{deltabar}) can only be realized
by possibility (b), the probability of which vanishes in the Grad limit.
Thus, the characteristic function $\chi$ is, in this limit,
\begin{equation}
\label{chardelbar}
\chi_{\overline{\Delta}_0({\bf y},{\bf v},t)}=
\left\{ \begin{array}{ll}
      0 & \mbox{if $0<t<T_0$}\\
      \chi_{\Gamma(t-T_0,t)} & \mbox{if $t>T_0$}\;.
      \end{array}
      \right.
\end{equation}
The characteristic function for the set $\Delta_0({\bf y},{\bf v},t)$ is
\begin{equation}
\label{chardel}
\chi_{\Delta_0({\bf y},{\bf v},t)}=1-
\chi_{\overline{\Delta}_0({\bf y},{\bf v},t)}\;.
\end{equation}

We now evaluate the characteristic function of the set $\Gamma(t_1,t_2)$.
Let $\cal L$ be the trajectory of the particle for $t_1<\tau<t_2$. Then,
\begin{equation}
\label{traj}
{\cal L}=\{{\bf x}(\tau) | t_1<\tau<t_2\}\;.
\end{equation}
The distance between $\cal L$ and a given point $\bf a$ is
\begin{equation}
\label{dist}
\mbox{dist}({\cal L},{\bf a})=\min_{t_1<\tau<t_2} \|{\bf x}(\tau)-{\bf a}\|\;.
\end{equation}
The characteristic function $\chi_{\Gamma(t_1,t_2)}$ --- for finite $N$
--- is given by
\begin{equation}
\label{chargamma}
\chi_{\Gamma(t_1,t_2)}
=\prod_{j=2}^N\theta[\mbox{dist}({\cal L},{\bf y}_j)-a]\;.
\end{equation}
The volume of the set $\Gamma(t_1,t_2)$ is then the integral
\begin{equation}
\label{volgamma}
\|\Gamma(t_1,t_2)\|=\frac{1}{A^{N-1}}\ \int_{\Omega^{N-1}} d{\bf Y}\
\chi_{\Gamma(t_1,t_2)}\approx \left[\frac{1}{A}\ 
\int_{\Omega} d{\bf y} \theta(\mbox{dist}({\cal L},{\bf y})-a)\right]^{N-1}\;.
\end{equation}
We have on the right-hand side of this expression assumed that we are close to
the Grad limit so that the positions of the scatterers become independent of 
each other.  We evaluate the integral 
\begin{equation}
\label{volgamma1}
\frac{1}{A}\ 
\int_{\Omega} d{\bf y} \theta(\mbox{dist}({\cal L},{\bf y})-a)=
\left[1-\frac{2aL(t_1,t_2)}{A}\right]+{\cal O}(a^2)\;,
\end{equation}
where $L(t_1,t_2)$ is the length of the curve $\cal L$.  
Combining (\ref{volgamma})
and (\ref{volgamma1}), and going to the Grad limit, we find that
\begin{equation}
\label{volgammagrad}
\|\Gamma(t_1,t_2)\|=e^{-2naL(t_1,t_2)}=e^{-\nu(t_2-t_1)}\;,
\end{equation}
where we have used that the particle moves with constant speed $v$ so that
$L(t_1,t_2)=v(t_2-t_1)$. 

Using Eq.\ (\ref{volgammagrad}) combined with Eqs.\ (\ref{chardelbar}) and 
(\ref{chardel}), we find that
\begin{equation}
\label{voldel0}
\|\Delta_0({\bf y},{\bf v},t)\|=
\left\{ \begin{array}{ll}
      1 & \mbox{if $0<t<T_0$}\\
      1-e^{-\nu T_0} & \mbox{if $t>T_0$}\;.
      \end{array}
      \right.
\end{equation}

We are now in the position to calculate the volume of 
$\Delta_k^{(0)}({\bf y},{\bf v}, t)$ through Eq.\ (\ref{deltaequal}) combined
with Eqs.\ (\ref{volgammagrad}) and (\ref{voldel0}).  

The set $\Gamma(t-kT_0,t)$, consists of all $\bf Y$
such that the particle has $k$ consecutive collisions with the scatterer at
$\bf y$ without collisions with other scatterers.  If we follow the
motion of the particle during the time interval\\ $[t-(k+1)T_0,t-kT_0]$ that
preceeds the time interval for which $\Gamma(t-kT_0,t)$ is defined, and
we assume $t$ large enough so that $t-(k+1)T_0>0$,
two histories are possible: (a) The particle has 
collided with other scatterers within the time interval 
$[t-(k+1)T_0,t-kT_0]$, or (b) it has collided with the scatterer at $\bf y$,
in which case it has {\it not\/} collided with any other scatterer during
this period. If (a) is the case, then the configuration belongs to the set
$\Delta_k^{(0)}({\bf y},{\bf v},t)$ --- see Eq.\ (\ref{deltaequal}) --- 
otherwise the configuration belongs to the set $\Gamma(t-(k+1)T_0,t)$.  Thus, 
we have that
\begin{equation}
\label{sectionunion}
\Gamma(t-kT_0,t)=\Delta_k^{(0)}({\bf y},{\bf v},t)\cup \Gamma(t-(k+1)T_0,t)\;.
\end{equation}
Furthermore, 
\begin{equation}
\label{empty}
\Delta_k^{(0)}({\bf y},{\bf v},t)\cap\Gamma(t-(k+1)T_0,t)=\emptyset\;.
\end{equation}
The volume of the set $\Delta_k^{(0)}({\bf y},{\bf v},t)$ is therefore
\begin{equation}
\label{volumdelta0}
\|\Delta_k^{(0)}({\bf y},{\bf v},t)\|=\|\Gamma(t-kT_0,t)\|-
\|\Gamma(t-(k+1)T_0,t)\|\;.
\end{equation}
Eq.\ (\ref{volgammagrad}), which is exact in the Grad limit, gives
\begin{equation}
\label{volumdelta0erlik}
\|\Delta_k^{(0)}({\bf y},{\bf v},t)\|=e^{-\nu kT_0}
\left(1-e^{-\nu T_0}\right)\qquad \mbox{if}\qquad t>(k+1)T_0\;.
\end{equation}
If we now assume that $kT_0<t<(k+1)T_0$ so that 
$\Gamma(t-(k+1)T_0,t)=\emptyset$, we have that 
$\Delta_k^{(0)}({\bf y},{\bf v},t)=\Gamma(t-kT_0,t)$ and consequently,
\begin{equation}
\label{verlikt}
\|\Delta_k^{(0)}({\bf y},{\bf v},t)\|=e^{-\nu kT_0} \qquad \mbox{if}\qquad
kT_0<t<(k+1)T_0\;.
\end{equation}

Using Eqs.\ (\ref{volumdelta0erlik}) and (\ref{verlikt}) (remember that
$S_0^{-k}$ depends on $\bf n$), we
may rewrite Eq.\ (\ref{ia2})
\begin{equation}
\label{ia3}
I_0\left({\bf y},{\bf v},{\bf n},t\right) 
=\left(1-e^{-\nu T_0}\right)\ \sum_{k=0}^{[t/T_0]-1} 
e^{-\nu kT_0} {\cal F}_k 
\left({\bf y},S_0^{-k}{\bf v},t-kT_0\right)
+e^{-\nu [t/T_0]T_0} {\cal F}_{[t/T_0]}
\left({\bf y},S_0^{-[t/T_0]}{\bf v},t-\left[\frac{t}{T_0}\right]T_0\right)\;.
\end{equation}

Our next task is to find the relation between $f_1({\bf x},{\bf v},t)$ and
${\cal F}_k({\bf y},{\bf v},t)$.
%------------------------------------------------------------------------------
\section{The Generalized Boltzmann Equation}
\label{genboltz}
The Grad limit of eq. (\ref{basic}) becomes
\begin{equation}
\label{asbasic}
\frac{D}{Dt}f_1({\bf x},{\bf v},t)=
an\ \int_{S^1} d{\bf n} ({\bf v}\cdot {\bf n}) 
[\theta({\bf v}\cdot {\bf n})b_{\bf n} +\theta(-{\bf v}\cdot {\bf n})]\
I_0({\bf x},{\bf v},{\bf n},t)\;,
\end{equation}
with $I_0$ given by eq. (\ref{ia3}) with ${\bf y}={\bf x}$.  

If $t<T_0$,
as a consequence of Eq.\ (\ref{ia3}) $I_0({\bf x},{\bf v},{\bf n},t)=
{\cal F}_0({\bf x},{\bf v},t)$.  To express ${\cal F}_0({\bf x},{\bf v},t)$
through $f_1({\bf x},{\bf v},t)$, we consider the right hand side of 
(\ref{gradf}) with $k=0$ and set for the moment
\begin{equation}
\label{ffn}
F({\bf x},{\bf v},t;{\bf y},{\bf Y})\equiv
F^{(N)}({\bf x},{\bf v},t;{\bf y},{\bf Y})
\end{equation}
to stress that it is a solution of the $N$-scatterer problem.  Near the Grad
limit, i.e., for sufficiently large $N$ and small $a$ provided 
$Na=\mbox{constant}$, we have 
\begin{equation}
\label{ffnt0}
F^{(N)}({\bf x},{\bf v},t=0;{\bf y},{\bf Y})\approx
\frac{1}{A}\ F^{(N-1)}({\bf x},{\bf v},t=0;{\bf Y})\;.
\end{equation}
(See Eq.\ (\ref{initialcond}) for $a\to 0$.)  Assuming that 
${\bf Y}\in \Delta_0({\bf y}+a{\bf n},{\bf v},t)$ (i.e., no collisions with
the scatterer at $\bf y$ before time $t$), we obtain from (\ref{formalsol})
\begin{equation}
\label{fnfn}
F^{(N)}({\bf y}+a{\bf n},{\bf v},t;{\bf y},{\bf Y})
=F^{(N)}(S^{(N-1)}_{-t} ({\bf y}+a{\bf n}),S^{(N-1)}_{-t}{\bf v},t=0;{\bf y}
{\bf Y})\;
\end{equation}
Combining the last two equalities, we get
\begin{equation}
\label{fnmore}
F^{(N)}({\bf y}+a{\bf n},{\bf v},t;{\bf y},{\bf Y})\approx
\frac{1}{A}\ F^{(N-1)}({\bf y}+a{\bf n},{\bf v},t;{\bf Y})\;.
\end{equation}
for ${\bf Y}\in \Delta_0({\bf y}+a{\bf n},{\bf v},t)$.  Thus, near the 
Grad limit we have from Eq.\ (\ref{gradf})
\begin{equation}
\label{gradf0}
{\cal F}_0({\bf y},{\bf v},t)\approx 
\frac{1}{\|\Delta_0({\bf y}+a{\bf n},{\bf v},t)\|}\ 
\int_{\Delta_0({\bf y}+a{\bf n},{\bf v},t)}
d{\bf Y} F^{(N-1)}({\bf y}+a{\bf n},{\bf v},t;{\bf Y})
\end{equation}
for all $t>0$.  
The simplest case $0<t<T_0$, leads to the approximate
equalities
\begin{equation}
\label{delta00}
\Delta_0({\bf y},{\bf v},t)\approx \Omega^{N-1}\;,
\end{equation}
\begin{equation}
\label{delta111}
\|\Delta_0({\bf y},{\bf v},t)\|\approx 1\quad \Rightarrow\quad
{\cal F}_0({\bf y},{\bf v},t)\approx f_1({\bf y},{\bf v},t)\;,
\end{equation}
which become exact in the Grad limit --- see Eq.\ (\ref{voldel0}).  
Hence, for $0<t<T_0$, Eq.\ 
(\ref{asbasic}) is identical to the usual Boltzmann equation
(\ref{normalbolt}) for $f_1({\bf x},{\bf v},t)$.  
We remark that this is perhaps the simplest way to extract the Boltzmann
equation directly from the Liouville equations in the Grad limit.

When $t>T_0$, $\Omega^N$ --- which is the space over which $F^{(N)}$ is
averaged in order to obtain $f_1$ --- is split into two disjoint sets,
\begin{equation}
\label{split}
\Omega^N={\cal A}_0\cup{\cal A}_1\;,
\end{equation}
where
\begin{equation}
\label{a0}
{\cal A}_0=\{({\bf y}_1,...,{\bf y}_N):\ \mbox{no collisions  
during the period $(t-T_0,t)$}\}\;.
\end{equation}
Both subsets ${\cal A}_0$ and ${\cal A}_1$ are defined for any given 
phase-time point $({\bf x},{\bf v},t)$, $t>T_0$.  If the particle is
not scattered during the time $(t-T_0,t)$, there was no earlier scattering
due to the periodicity of the motion.  Thus, the subsets 
${\cal A}_0$ and ${\cal A}_1$ do not actually depend on time $t>T_0$ and
can be equally well defined at $t=T_0$.  The solution of the Liouville
equation (\ref{liouville}) with initial conditions (\ref{initialcond})
for $({\bf y}_1,{\bf Y})\in {\cal A}_0$ is 
\begin{equation}
\label{initialcondmore}
F^{(N)}({\bf x},{\bf v},t;{\bf y}_1,...,{\bf y}_N)
=f_1(S_0(-t){\bf x},S_0(-t){\bf v},0)\left[\prod_{i=1}^N \ 
\theta(|S_0(-t){\bf x}-{\bf y}_i|-a)\right]\ 
\frac{\rho({\bf y}_1,...,{\bf y}_N)}{1-\pi n a^2}\;,
\end{equation}
where $S_0(\Delta t)$ is the shift operator along the 
collisionless trajectory of the particle from $t$ to $t+\Delta t$.
Using Eq.\ (\ref{volgammagrad}), we find that in the Grad limit
\begin{equation}
\label{vola0}
\|{\cal A}_0\|=e^{-\nu T_0}\;,
\end{equation}
and consequently
\begin{equation}
\label{vola1}
\|{\cal A}_1\|=1-e^{-\nu T_0}\;.
\end{equation}
We may write the distribution $f_1({\bf x},{\bf v},t)$ for $t>T_0$ as the
sum of $F$ averaged over ${\cal A}_0$ and ${\cal A}_1$,
\begin{equation}
\label{faveraged}
f_1({\bf x},{\bf v},t)=\|{\cal A}_0\| f_1^{(0)}({\bf x},{\bf v},t)
+\|{\cal A}_1\| f_1^{(1)}({\bf x},{\bf v},t)\;,
\end{equation}
where
\begin{equation}
\label{fa0}
f_1^{(i)}({\bf x},{\bf v},t)=\frac{1}{\|{\cal A}_i\|}
\int_{{\cal A}_i} d{\bf y}
d{\bf Y}\ F^{(N)}({\bf x},{\bf v},t;{\bf y},{\bf Y})\;,\quad i=0,1\;.
\end{equation}
We generalize the notation (\ref{faveraged}) to $0<t<T_0$ by setting 
${\cal A}_0=\emptyset$ and ${\cal A}_1=\Omega^N$ in such a case.

Let us consider Eqs.\ (\ref{yin0}), (\ref{gradf}), (\ref{asbasic}) and 
(\ref{ia3}) for
$t>T_0$.  If $k\ge 1$, in (\ref{gradf}), we may repeat the argumentation
leading to equality (\ref{gradf0}) and obtain 
\begin{equation}
\label{gradf1}
{\cal F}_k({\bf y},S_0^{-k}{\bf v},t-kT_0)\approx 
\frac{1}{\|\Delta_k^{(0)}({\bf y}+aS_0^{-k}{\bf n},{\bf v},t)\|}\ 
\int_{\Delta_k^{(0)}(S_0^{-k}({\bf y}+a{\bf n}),{\bf v},t)}
d{\bf Y} F^{(N-1)}(S_0^{-k}({\bf y}+a{\bf n}),S_0^{-k}{\bf v},t-kT_0;{\bf Y})\;,
\end{equation}
where $k=0,...,[t/T_0]$, near the Grad limit.  A minor change of notation (set $N=\tilde N+1$ and
${\bf y}_i=\tilde{\bf y}_{i-1}$, $i=2,...$) leads to formulas similar to 
the definition (\ref{fa0}) of $f_1^{(1)}$ with the only difference that
the function $F$ is averaged not over the whole ``ergodic" domain 
${\cal A}_1$, but over a subdomain $\Delta_k^{(0)}\subset {\cal A}_1$.
The final step is to assume that in the Grad limit
\begin{equation}
\label{gradfk}
{\cal F}_k({\bf y},{\bf v},t)
=f_1^{(1)}({\bf y},{\bf v},t)\;,\quad k=0,1...\;,\quad t>0\;.
\end{equation}
This assumption means that we can neglect in this limit the difference
between average values of $F({\bf x},{\bf v},t;{\bf y}_1,...{\bf y}_N)$
taken over different subsets $\Delta_k^{(0)}$ of the ``ergodic" set 
${\cal A}_1$.  This
assumption is in fact one of the central postulates of kinetic theory
(the average behavior of the test particle is the same for almost all
configurations of scatterers) and we admit it here without proof.
The corresponding rigorous result for the linear Boltzmann equation 
(without magnetic field) was obtained in \cite{BBS83}.  Finally, we note
that the representation of $f_1({\bf x},{\bf v},t)$ in the form 
(\ref{faveraged}) (cycling and wandering particles, see Refs.\ 
\cite{BMHH95,BMHH97}) makes sense only for $t>T_0$.  At the first stage
of the motion, $0<t<T_0$, all particles should be considered as wandering,
i.e., having a chance to undergo a collision in the future.  This explains
why we put ${\cal A}_1=\Omega^N$ and therefore $f_1^{(1)}\equiv f_1$ for
$0<t<T_0$, see comment after Eq.\ (\ref{fa0}). 
The above formulas (\ref{asbasic}), (\ref{initialcondmore}) ---
(\ref{fa0}) define completely the non-markovian kinetic equation 
(\ref{asbasic}) in the Grad limit.  The final step is to reduce the
equation to a more convenient form.  To this end we define $f^G$ by
Eq.\ (\ref{fgdef}).
Thus, we have the normalization
\begin{equation}
\label{normalization}
\int d{\bf x}\ d{\bf v}\ f^G({\bf x},{\bf v},t)=
\left\{ \begin{array}{ll} 
      1                & \mbox{if $0<t<T_0$}\\
      1-e^{-\nu T_0} & \mbox{if $t>T_0$}\;.
      \end{array}
      \right.
\end{equation}
Eq.\ (\ref{ia3}) may thus be written
\begin{equation}
\label{ia8}
I_0\left({\bf x},{\bf v},t\right) 
=\sum_{k=0}^{[t/T_0]} 
e^{-\nu kT_0} f^G\left({\bf y},S_0^{-k}{\bf v},t-kT_0\right)\;.
\end{equation}

The ``non-colliding" distribution function $f_1^{(0)}$ satisfies the 
transport equation
\begin{equation}
\label{f0eq}
\frac{D}{Dt}\ f_1^{(0)}({\bf x},{\bf v},t)=0\;.
\end{equation}
The ``colliding" distribution function, $f_1^{(1)}$, satisfies the equation
\begin{equation}
\label{f1eq}
\frac{D}{Dt}\ \left(1-e^{-\nu T_0}\right)\ f_1^{(1)}({\bf x},{\bf v},t)=
n\ \int_{S^1} d{\bf n} ({\bf v}\cdot {\bf n})
[\theta({\bf v}\cdot {\bf n})b_{\bf n} +\theta(-{\bf v}\cdot {\bf n})]\
\left[\sum_{k=0}^{[t/T_0]} 
e^{-\nu kT_0} f^G\left({\bf y},S_0^{-k}{\bf v},t-kT_0\right)\right]\;.
\end{equation}
When $t>T_0$, we use the definition (\ref{fgdef}) and write this equation
\begin{equation}
\label{fgeqe}
\frac{D}{Dt}\ \ f^G({\bf x},{\bf v},t)=
n\ \sum_{k=0}^{[t/T_0]}\ e^{-\nu kT_0}\ 
\int_{S^1} d{\bf n} ({\bf v}\cdot {\bf n})
[\theta({\bf v}\cdot {\bf n})b_{\bf n} +\theta(-{\bf v}\cdot {\bf n})]\
f^G\left({\bf y},S_0^{-k}{\bf v},t-kT_0\right)\;.
\end{equation}
When $t<T_0$, we add Eqs.\ (\ref{f0eq}) and (\ref{f1eq}).  The resulting 
equation has the same form as Eq.\ (\ref{fgeqe}) --- which is the generalized
Boltzmann equation of Ref.\ \cite{BMHH95,BMHH97}.
%--------------------------------------------------------------------
\section{Conclusion}
\label{conclusion}

We have presented a systematic derivation of the generalized Boltzmann equation
(\ref{fg1}) from the Liouville equation in the Grad limit.  The derivation
was done at the traditional physicist level of rigor and we stressed several
points which were not proved mathematically.  Moreover, a convergence of the
right hand side of Eq.\ (\ref{hier5}) for $f_1({\bf x},{\bf v},t)$ is 
obviously not equivalent to convergence of the ``true" function $f_1^{(N)}$
to the solution of the generalized Boltzmann equation.  Therefore, there still
remains a considerable effort from a mathematical point of view to
prove the validity of the generalized Boltzmann equation at the same level
of rigor as it was done for the
usual Boltzmann equation (see Ref.\ \cite{HS} and references therein).
However, some of the basic ideas needed and some hints at the
difficulties that need to be overcome to construct such a proof can be
found in the above ``physical derivation."

Another open problem is to clarify the connection of the generalized 
Boltzmann equation with the BBGKY hierarchy.  The arguments discussed
above show that the validity of the generalized Boltzmann equation follows
directly from $N$-scatterer dynamics in the Grad limit.  The generalized
Boltzmann equation is clearly the non-markovian equation for a system with
{\it strong\/} long-time pair correlations in contradistinction to the
usual Boltzmann equation.  It is therefore important to understand what
the correlation functions look like in this case.  

In order to do this, one needs to consider the whole hierarchy and find its
asymptotic solution in the Grad limit.  We hope to do this in a forthcoming
paper.
\bigskip\bigskip

A.\ V.\ B.\ thanks the Norwegian Research Council for financial support.
A.\ H.\ thanks the CNRS for support during his stay in Lyon. J.\ P.\ thanks 
the KBN (Committee for Scientific Research, Poland) for support through grant 
2 P03B 03512. J.\ P.\ furthermore thanks ENSL (Lyon) and NTNU (Trondheim)
for hospitality.
%--------------------------------------------------------------------
\section{Appendix}

In this Appendix, we construct the shift operator $S_a$ which was
defined in (\ref{yin0}).  It shifts the particle from a position right before
a collision with the scatterer at $\bf y$ along a collisionless trajectory
to a new precollisional position on the surface of the scatterer.

We consider in the following a succession of collisions between the charged 
particle and a single scatterer.  Assume that the particle follows the 
trajectory marked ``1" in Fig.\ \ref{fig1}. This trajectory is a circle arc 
centered at a distance $\Delta$ from the center of the scatterer.  The 
scattering occurs on the surface of the disk, with a scattering angle $\psi$, 
and the particle shifts to a new cyclotron orbit ``2". Clearly 
$-({\bf v}'\cdot {\bf n}) =({\bf v}\cdot{\bf n})=|{\bf v}|\sin(\psi/2)$, where 
${\bf v}'$ is the precollisional velocity and $\bf v$ the postcollisional 
one. As a result of this symmetry, the 
center of the new orbit ``2" has the same distance $\Delta$ from 
the center of the scatterer as the previous orbit. Orbit ``2" leads back to the
surface of the scatterer where a new collision occurs, and the particle is 
shifted to a third cyclotron orbit ``3".  The direction of ${\bf n}$ in
the first collision and $\bf n$ of the second collision is shifted by an
angle $2\beta$, where $\beta$ is given by
\begin{equation}
\label{pytagoras}
\cos\beta=\frac{\Delta^2-R^2+a^2}{2a\Delta}\;.
\end{equation}
Here $R$ is the cyclotron radius, and $a$ is the radius of the scatterer.  

The angle between two subsequent points of collision is $2\beta$ as
seen from the center of the scatterer. Similarly, the angle between
these two points, as seen from the center of the cyclotron orbit one to the
other, is $2\gamma$, with
\begin{equation}   \label{pyt}
  \cos \gamma =\frac{R^2+\Delta^2-a^2}{2R\Delta}.
\end{equation}
The scattering angle in the second
collision remains equal to that of the first, $\psi$. Note that in the Grad 
limit, $\beta = (\pi-\psi)/2$ can take any
value from 0 to $\pi /2$. On the other hand, $\gamma \to 0$ for 
{\it all} collisions in the Grad limit.

Thus, if we set ${\bf v}=(v,\varphi_v)$ and ${\bf n}=(1,\varphi_n)$ in 
polar coordinates, the shift operator $S_a$ is, with reference to
the figure,
\begin{equation}
\label{sv}
S_a\ (v,\varphi_v)=(v,\varphi_v+\psi -2\gamma) 
 \stackrel{\rm Grad}{\to} (v,\varphi_v +\psi)\;,
\end{equation}
with $\gamma$ given by (\ref{pyt}). Similarly,
\begin{equation}
\label{sn}
S_a\ (1,\varphi_n)=(1,\varphi_n+2\beta)\;,
\end{equation}
where $\beta$ is given by (\ref{pytagoras}).

In polar coordinates, the generalized Boltzmann equation (\ref{fg1}) becomes
\begin{equation}
\label{fg1phi}
\frac{D}{Dt}\ \ f^G({\bf x},\varphi_v,t)=
\frac{na}{4}\ \sum_{k=0}^{[t/T_0]}\ e^{-\nu kT_0}\ 
\int_{-\pi}^{\pi} d\psi \sin\left|\frac{\psi}{2}\right|\ \left[
f^G({\bf x},\varphi_v-(k+1)\psi,t-kT_0)
-f^G({\bf x},\varphi_v-k\psi,t-kT_0)\right]\;.
\end{equation}
This was the form of the generalized Boltzmann equation  used in
Refs.\ \cite{BMHH95,BMHH97}.

% --------------------------------------------------------------------
% BIBLIOGRAPHY
% --------------------------------------------------------------------

% --------------------------------------------------------------------
% FIGURE CAPTIONS
% --------------------------------------------------------------------
\begin{figure}
\caption[x]{The charged particle follows orbit ``1" and collides with 
the hard disk, thus shifting to orbit ``2".  It then recollides with 
the disk and shifts to orbit ``3".  The cyclotron radius is $R$ and the 
distance between the centers of the cyclotron orbits to the center of the
scatterer is $\Delta$.  The angles separating two subsequent collisions
is $2\beta$.
\label{fig1}
}
\end{figure}
% --------------------------------------------------------------------
\end{document}